\documentclass[a4paper]{article}

\usepackage{INTERSPEECH2020}
\usepackage{multirow}
\usepackage{comment}
\usepackage{xcolor}

\title{Attention and Encoder-Decoder based models for transforming articulatory movements at different speaking rates}
\name{Abhayjeet Singh,  Aravind Illa, Prasanta Kumar Ghosh}
\address{Electrical Engineering, Indian Institute of Science (IISc), Bangalore-560012, India}
\email{spirelab.ee@iisc.ac.in}

\begin{document}

\maketitle
\begin{abstract}
While speaking at different rates, articulators (like tongue, lips) tend to move differently and the enunciations are also of different durations. In the past, affine transformation  and DNN have been used to transform articulatory movements from neutral to  fast(N2F) and neutral to slow(N2S) speaking rates \cite{astha}. In this work, we  improve over the existing transformation techniques by modeling rate specific durations  and their transformation using AstNet, an encoder-decoder framework with attention. In the  current work, we propose an encoder-decoder architecture using LSTMs which  generates smoother predicted articulatory trajectories. For modeling duration variations  across speaking rates, we deploy attention network, which eliminates the need to align  trajectories in different rates using DTW. We perform a phoneme  specific duration analysis to examine how well duration is transformed using the  proposed AstNet. As the range of articulatory motions is correlated with speaking rate,  we also analyze amplitude of the transformed articulatory movements at different rates  compared to their original counterparts, to examine how well the proposed AstNet  predicts the extent of articulatory movements in N2F and N2S. We observe that AstNet could model both duration and extent of  articulatory movements better than the existing transformation techniques resulting in more accurate transformed articulatory trajectories.
\end{abstract}

\section{Introduction}
Speech production involves planning and coordination of different articulators including, lips, jaw, tongue and velum \cite{goldstein2003articulatory}. Apart form phonetic information, various para-linguistic factors could influence the articulatory movements, and in turn, acoustics. One such factor is speaking rate,
 which is defined as the number of phonemes spoken per second \cite{phonetics1951study}. Speaking rate affects acoustic properties  such as vowel duration \cite{kuwabara1997acoustic}, vowel formant frequencies \cite{lindblom1963spectrographic,gay1978effect,moon1994interaction}, consonant-vowel co-articulation \cite{agwuele2008effect}, average syllable duration \cite{miller1984articulation} and pronunciation \cite{fosler1999effects} due to changes in the articulation \cite{ILLA202075,kuberski2019speed}. Variation in acoustic features because of changing speaking rate directly impacts the performance of several speech applications including automatic speech recognition (ASR) \cite{benzeghiba2007automatic,meyer2011effect,stern1996signal} that are typically designed for speech characterized by the average speaking rate. Previous studies also reported speaking rate specific changes in articulation such as the rate of articulation, range of articulatory movements and the degree of co-articulation \cite{berry2011speaking}. Understanding the nature of articulatory dynamics at different speaking rates and their inter relationships could help in developing speech based systems that are robust to variations in speaking rate.

In this work, we learn a transformation function which maps the articulatory movements from neutral to fast speaking rate (N2F), and also for neutral to slow speaking rate (N2S) using an encoder-decoder framework with attention network. 
The transformation should be able to learn the variations in duration and range of movements of articulators from source speaking rate to target speaking rate.
Preliminary work has been done on this topic \cite{astha}, where variations in the range of articulatory movements are modeled using various transformation methods like an affine transformation with both diagonal and full matrix and a non-linear transformation modeled by a deep neural network (DNN). As all these transformation methods learn work at the frame level, which require dynamic time warping (DTW) to time-align articulatory movement trajectories at different rates. To learn the optimal transformation function (TF), an iterative approach was followed to optimize TF and the optimal warping path till convergence is achieved. The limitations of this approach lies in modelling the duration variations across speaking rates.
 To overcome these shortcomings in \cite{astha}, the proposed approach utilizes attention mechanism to capture the duration variations to improve the performance. The proposed encoder-decoder based attention network used to perform Articulatory trajectories Speaking rate Transformation is denoted as AstNet. AstNet is an end-to-end model which takes articulatory movements at one rate as input and generates at a different rate. This is in contrast with the iterative approach followed in \cite{astha}. In addition, we don't need to time-align prior to training the model as the attention network learns the duration modelling and by using ``stop token" method, end of sequence is predicted at the time of inference. We use long-short-term memory (LSTM) units in both encoder and decoder, and this results in preserving the smoothness characteristics in the predicted articulatory movements.    

The experimental setup and data-set for this work are similar to the preliminary work \cite{astha}. 
The proposed AstNet model outperforms the technique proposed in \cite{astha} in both cases (N2F and N2S) for all subjects. DTW distance is used as an evaluation metric, with lower distances indicating better performance. As it was reported that different sound units are affected differently in different speaking rates \cite{kuwabara1997acoustic}, we perform analysis on the duration of phonemes at the source (neutral) rate, ground-truth duration at the target rate and the same at target rate following articulatory movement prediction.
We also analyze the range of articulatory movements by computing  standard deviation of articulatory trajectory (SDAT) \cite{ILLA202075} at source (neutral) rate as well as original and predicted trajectories at target rate. These analysis on duration and range of articulatory movements could help to understand the modeling capabilities of AstNet.

\section{Dataset}
\begin{figure}[h]
   \centering
    \vspace{-.2cm}
   \centerline{\includegraphics[trim = 0mm 0mm 0mm 0mm, clip, width=7.5cm]{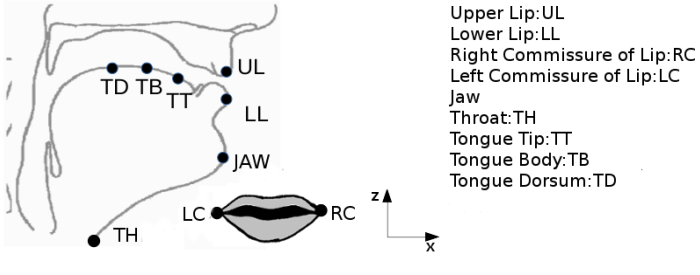}} 
    \vspace{-.2cm}
   	\caption{Schematic diagram indicating the placement of EMA sensors \cite{illa2018low}.}
   	\label{EmaSetUp}
 \vspace{-.5cm}
 \end{figure}

In this study, we collected articulatory data for 460 sentences spoken at three different rates: neutral, fast and slow. This data was collected from 5 subjects out of which three were male (M1, M2, M3) and two were female (F1, F2) of age 19, 22, 24, 28 and 22 respectively. All subjects are native Indians reported to have no speech disorders. We took 460 phonetically balanced sentences from MOCHA-TIMIT corpus \cite{timit}. The phonetic transcriptions for these sentences were obtained using SONIC speech recognizer \cite{pellom2001sonic}, which uses 55 phoneme set of American English in its lexicon \footnote{https://www.yumpu.com/en/document/read/9555742/sonic-the-university-of-colorado-continuous-speech-recognizer}. Among these, five phonemes (ls, ga, SIL, br and lg) did not appear in the force-alignment output. In this study, we ignore the speaker specific phoneme variations. Figure \ref{EmaSetUp} shows the placement \cite{optimal} of nine articulators among which seven are along the mid-sagittal plane (indicated by X and Z directions) and the remaining two (RC and LC) are in frontal plane, to record articulatory movements using electromagnetic articulograph AG501 \cite{AG501}. We obtain 18 articulatory trajectories corresponding to Upper Lip (ULx, ULz), Lower Lip (LLx, LLz), Right Commissure of Lip (RCx, RCz), Left Commissure of Lip (LCx, LCz), Jaw (JAWx, JAWz), Throat (THx, THz), Tongue Tip (TTx, TTz), Tongue Body (TBx, TBz), Tongue Dorsum (TDx, TDz), for each utterance. Articulatory movements are recorded at sampling rate of 250 Hz. It is known that the articulatory movements are low pass in nature \cite{ghosh2010generalized}, therefore we first low pass filtered with a cutoff frequency 25 Hz to avoid high frequency noise incurred in EMA data acquisition. These are further down-sampled from 250 Hz to 100 Hz. 
We also recorded synchronous acoustic of speech using a t.bone EM9600 shotgun \cite{EMG9600}, unidirectional electret condenser microphone.
 
The recording of the sentences in the three speaking rates for each subject was held in three different sessions. In the first session, the subject was asked to speak in his/her normal speaking rate, from which, after silence removal, their neutral speaking (phone) rate was computed. In the subsequent sessions the subjects were required to speak at 2 times their neutral speaking rate to record fast speech and reduce their speaking rate by half to record slow speech. 
For this work, we utilized the same dataset used in our previous work \cite{astha}, where more details regarding the dataset can be found.

\section{AstNet: Transforming articulatory movements to slow/fast speaking rate}
Mapping between articulatory trajectories spoken at different rates is challenging, as the duration of an articulatory trajectory is inversely proportional to its speaking rate and different phoneme durations vary differently with speaking rates \cite{MUTO2005277}. To model duration variations while transforming speaking rate, we propose AstNet, an encoder-decoder architecture with a location sensitive attention to learn alignments between sequences of varying durations.

\begin{figure}[h]
   \centering
    \vspace{-.5cm}
   \centerline{\includegraphics[trim = 0mm 0mm 0mm 0mm, clip, width=8cm,height=7cm]{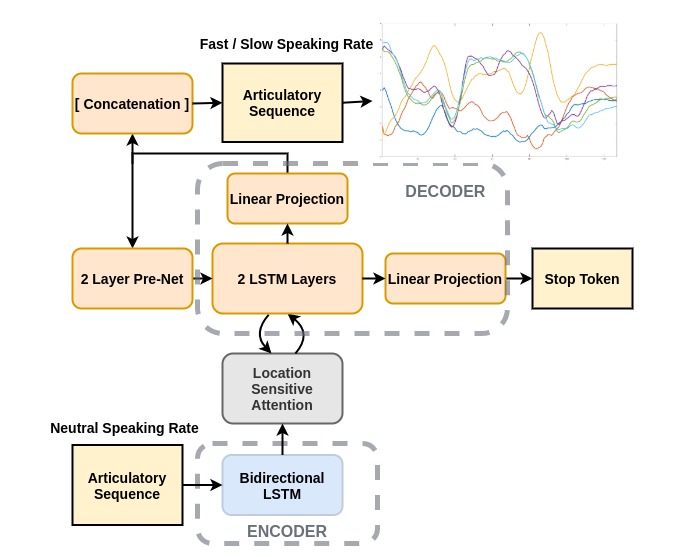}} 
    \vspace{-.4cm}
   	\caption{Block diagram representing the AstNet architecture}
   	\label{BD}
   	\vspace{-0.7cm}
 \end{figure}
 
Figure \ref{BD}, illustrates the AstNet architecture, which consists of encoder, attention and decoder. The encoder comprises a single BLSTM layer with 512 hidden units, which takes an articulatory sequence at a specific rate ($rate_{inp}$) as input $S_1=\{s_1,s_2,..s_{N}\}$ and maps it to the hidden states $E=\{e_1,e_2,..e_N\}$, which acts as an input to the attention. The attention network models the time alignment between the encoder and decoder hidden states. The decoder hidden states, $D=\{d_1,d_2,..d_M\}$, are utilized to generate the articulatory movement trajectories at a specific target speaking rate ($rate_{out}$). Attention network is a location-sensitive attention network which iterates over previous decoder output ($d_{t-1}$) and attention weights ($\alpha_{n,t-1}$) and all the encoder hidden states ($E$). The attention mechanism is governed by the equations below \cite{ASRattention}:
\begin{equation}
 \alpha_{n,t}=\sigma (\text{score}(d_{t-1},\alpha_{n,t-1},E))
\end{equation}
 \begin{equation}
 \text{score:}\ s_{n,t}=w^T \text{tanh}(Wd_{t-1}+Ve_n+Uf_{j,t}+b)
\end{equation} 
\begin{equation}
    \text{Context vector: } g_t=\sum_{j=1}^N\alpha_{j,t}e_{j}
\end{equation}
where $\alpha_{n,t}$ are \textit{attention weights} and $n \epsilon \{1,2..N\}$ and parameters of the attention network are denoted by $W$, $V$, $U$ weight matrices and $w$, $b$ denote the projection and bias vectors, respectively. In Eq.(2), $f_{j,t}$ is computed by $f_t=F*\alpha_{t-1}$, which incorporates location  awareness to the attention network \cite{ASRattention}, where $F$ is a weight matrix and $\alpha_{t-1}$ is the set of previous time-step alignments.
In Eq.(2), attention scores are computed as a function of encoder outputs ($e_n$) and previous attention weights ($\alpha_{n,t-1}$) and decoder output ($d_{t-1}$), which are further normalized using \textit{softmax} to obtain attention weights. These obtained attention weights are utilized to compute fixed context vector as described by Eq.(3).

The decoder consists of two uni-directional LSTM layers with 1024 units, followed by a linear projection layer to predict the articulatory movements  at $rate_{out}$. The decoder computes the final output ( $d_t \sim Decoder(d_{t-1},g_{t})$) 
from the previous state output ($d_{t-1}$) and attention context vector ($g_t$). 
To compute $d_{t-1}$, decoder's previous output is transformed by a two layered fully-connected network with 256 units (Pre-Net). The decoder hidden state outputs are further projected using two linear layers, one for articulatory sequence prediction and other to predict end of the sequence. For end sequence prediction, the decoder LSTM output and attention context are concatenated and projected down to a scalar and then passed through sigmoid activation to predict the probability that the output sequence has completed. This ``Stop Token" prediction is used during inference to allow the model to dynamically determine when to terminate generation instead of always generating for a fixed duration. We vary the number of maximum steps taken by decoder according to $rate_{out}$ articulatory sequences' maximum duration. This hyper parameter comes into play only at the time of inference, to bound the upper limit of the predicted sequence\textquotesingle s duration.

\section{Experimental Setup}
In this work, we perform two different rate transformations using the AstNet, the first one is the neutral rate to fast rate transformation of articulatory movements (N2F) and the other one is the neutral rate to slow rate articulatory movement transformation (N2S). In both the cases N2F and N2S, we use a four fold setup. We divide the data into training and test set in the ratio of 3:1. Before training, articulatory trajectories are first made zero-mean and then transformations are learnt separately for N2F and N2S cases in each of the four folds from five subjects.

\textit{Training Approach:} Typically neural network models demand large amount of training data to achieve better performance. To overcome the scarcity of articulatory data to train a subject specific model, we deploy the training approach proposed in \cite{illa2018low}. At the first step, we pool the training data from all the subjects and train a generic model to learn the mapping from an articulatory trajectory at $rate_{inp}$ to an articulatory trajectory at $rate_{out}$. This is named as Generalized or Generic training. In the second stage, we fine tune the generic model weights with respect to the target speaker, to learn speaker specific models. We also train subject-dependent models for each subject.

\textit{Evaluation procedure:} For a given scheme, we indicate $D_T = \{d_1,d_2,d_3,d_4\}$, where $d_i \in \Bbb R ^{1\times115}$ consists of the DTW distances between the articulatory trajectory at $rate_{out}$ and the corresponding ground-truth trajectory for the 115 test utterances in the $i^{th}$ fold. In order to evaluate the performances of different schemes, we report the average and standard deviation of $D_T$ for every subject. Lower the average $D_T$ for a given scheme, better is the performance of the model. Therefore, the best scheme is the one which results in the least average $D_T$.
\begin{table}[th]
  \footnotesize
  \caption{Average (standard deviation) of $D_T$ (in mm) for N2F transformation.}
  \label{results_fast}
  \centering
  \begin{tabular}{p{5em} c c c c c c}
    \toprule
    \textbf{Model}&\textbf{F1}&\textbf{F2}&\textbf{M1}&\textbf{M2}&\textbf{M3} \\
    \midrule
    \multirow{2}{5em}{\textbf{IT-DTW}} & 6.51 & 6.60 & 6.71 & 6.00 & 5.64 \\
    & (0.85) & (1.09) & (1.18) & (0.93) & (1.04) \\
    \multirow{2}{5em}{\textbf{Baseline [1]}} & 4.79 & 5.04 & 4.72 & 4.86 & 4.58 \\
    & (0.63) & (0.90) & (0.90) & (0.76) & (0.90) \\
    \multirow{2}{5em}{\textbf{AstNet (Subj Dep)}} & 5.21 & 7.65 & 5.57 & 5.83 & 4.74 \\
    & (1.17) & (1.49) & (1.46) & (1.57) & (1.22) \\
    \multirow{2}{5em}{\textbf{AstNet (Generic)}} & 4.92 & 5.18 & 4.82 & 4.83 & 4.63 \\
    & (0.60) & (0.78) & (0.90) & (0.73) & (0.77) \\
    \multirow{2}{5em}{\textbf{AstNet (finetuned)}} & \textbf{4.69} & \textbf{4.95} & \textbf{4.59} & \textbf{4.65} & \textbf{4.45} \\
    & (0.57) & (0.76) & (0.84) & (0.74) & (0.78) \\
    \bottomrule
  \end{tabular}
  \vspace{-0.4cm}
\end{table}

\begin{table}[th]
  \footnotesize
  \caption{Average (standard deviation) of $D_T$ (in mm) for N2S transformation}
  \label{results_slow}
  \centering
  \begin{tabular}{p{5em} c c c c c c}
    \toprule
    \textbf{Model} & \textbf{F1} & \textbf{F2} & \textbf{M1} & \textbf{M2} & \textbf{M3} \\
    \midrule
    \multirow{2}{5em}{\textbf{IT-DTW}} & 6.46 & 8.27 & 7.32 & 6.03 & 6.30 \\
    & (0.90) & (1.27) & (1.07) & (0.77) & (0.89) \\
    \multirow{2}{5em}{\textbf{Baseline [1]}} & 5.05 & 7.37 & 6.30 & 5.35 & 5.52 \\
    & (0.89) & (1.11) & (1.03) & (0.75) & (0.80) \\
    \multirow{2}{5em}{\textbf{AstNet (Subj Dep)}} & 7.86 & 12.70 & 9.67 & 9.18 & 8.29 \\
    & (1.10) & (1.53) & (1.36) & (1.02) & (1.74) \\
    \multirow{2}{5em}{\textbf{AstNet (Generic)}} & 5.80 & 8.33 & 7.44 & 5.89 & 5.98 \\
    & (0.96) & (1.31) & (1.31) & (0.82) & (0.95) \\
    \multirow{2}{5em}{\textbf{AstNet (finetuned)}} & \textbf{4.76} & \textbf{6.83} & \textbf{6.10} & \textbf{5.01} & \textbf{4.95} \\
    & (0.83) & (1.14) & (1.27) & (0.89) & (0.92) \\
    \bottomrule
  \end{tabular}
  \vspace{-0.45cm}
\end{table}
\section{Results and Discussion}
First we compare the performance of AstNet with baseline approach, then we perform analysis on the transformed articulatory trajectories to verify speaking rate specific characteristics.

\subsection{Performance of AstNet:}
Estimation of articulatory movement at different speaking rates has been done in \cite{astha}, where they consider learning one-to-one mapping between articulatory movements at different rates using three schemes: 1. Full affine transformation matrix, 2. Diagonal affine transformation matrix and 3. Non-Linear transformation function. To learn one-to-one mapping they align the articulatory movements at different rates using Dynamic Time Warping (DTW), so that all corresponding articulatory trajectories at $rate_{inp}$ and $rate_{out}$ are of equal durations. We consider the best results from \cite{astha} as our baseline i.e., the lowest possible average DTW values for all five subjects, and compare the results of AstNet. Table \ref{results_fast} reports the DTW values of different approaches for N2F transformation.
In first row we report DTW distance between original neutral and fast speaking rate, as a naive baseline with identity transformation (IT-DTW), followed by DTW values reported in [1] as a baseline. The rest of the values in Table \ref{results_fast} from third to the last rows correspond to AstNet with different training schemes.  Similarly, for N2S transformation we report DTW values of respective models in Table \ref{results_slow}.
\begin{figure}[h]
   \centering
    \vspace{-.3cm}
   \centerline{\includegraphics[trim = 100mm 180mm 100mm 60mm, clip, width=9.8cm,height=6.8cm]{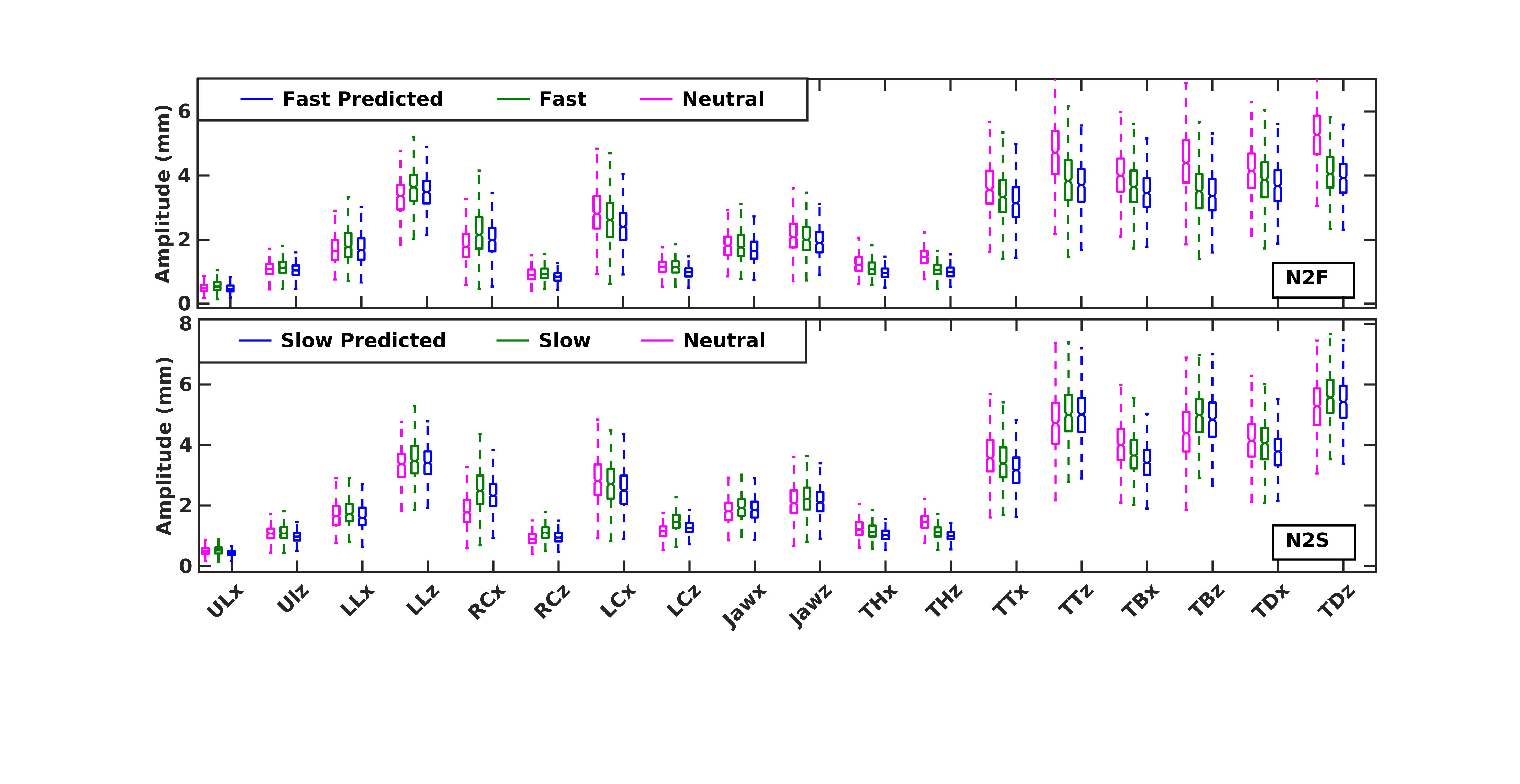}} 
    \vspace{-.5cm}
   	\caption{Standard deviation of articulatory trajectories (SDAT) over 18 articulators (N2F and N2S)}
   	\label{amp analysis}
 \vspace{-.3cm}
 \end{figure}

 \begin{figure*}[]
   \centering
   \centerline{\includegraphics[trim = 60mm 100mm 60mm 22mm, clip, width=\textwidth]{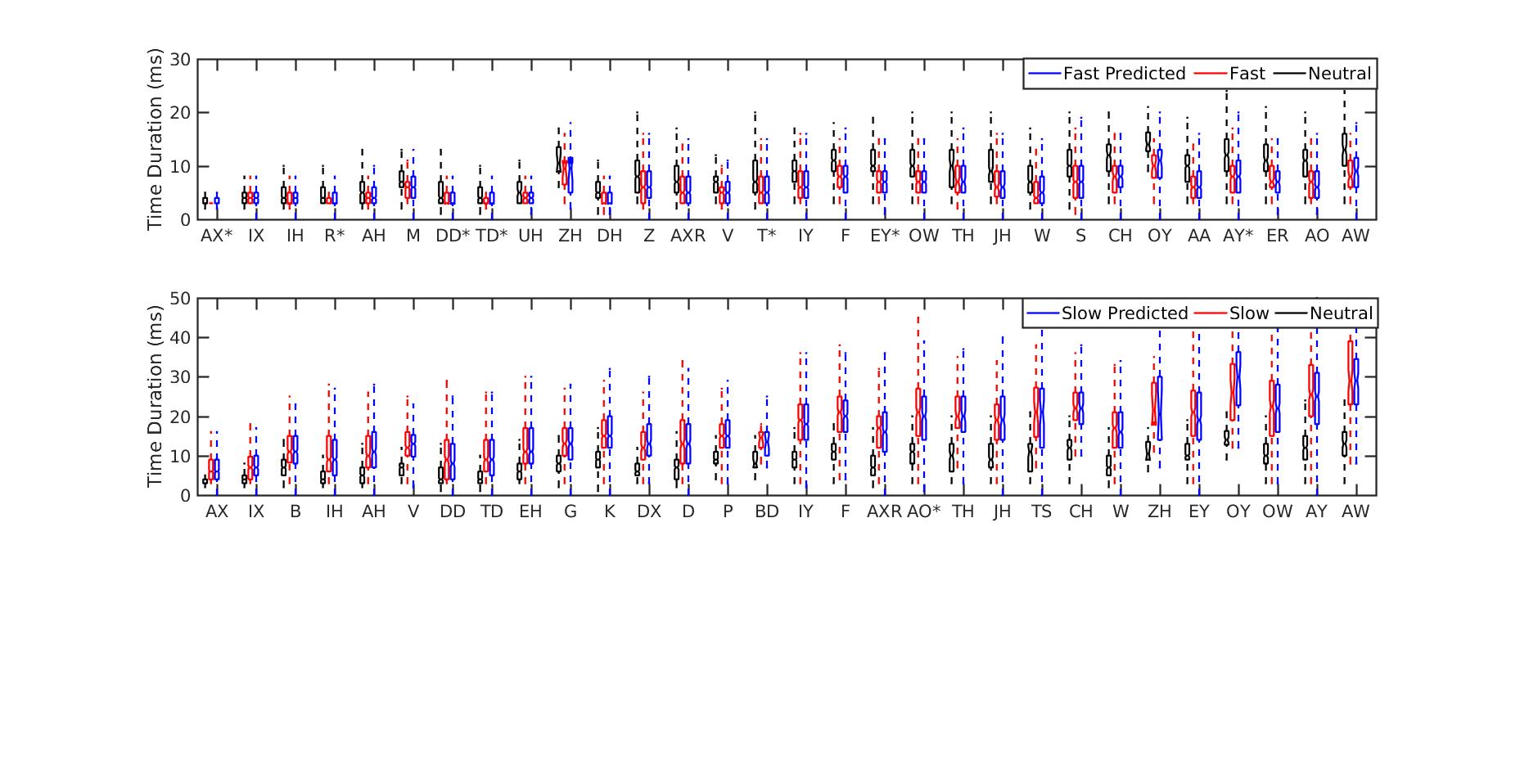}} 
    \vspace{-.8cm}
   	\caption{Duration analysis on a 50 phoneme set for N2F and N2S transformation (only 30 phonemes are shown).}
   	\label{duration_analysis_n2f}
   	\vspace{-0.5cm}
 \end{figure*}
 
In both the tables subject-dependent models have the lowest performance (highest DTW distance) among different AstNet models for all subjects. It is also to be noted that the DTW distances are poor when compared with IT-DTW. This is due to the lack of training data for each subject's model. In case of generic models in which we train a single model for all subjects, the performance is better than subject-dependent models and IT-DTW. Generic models learn the transformations for N2F and N2S cases which are common for all subjects. We further train this generic  model over each subject separately, giving us fine-tuned models, one for each subject. The finetuned AstNet performs better among all models including the baseline [1], as fine-tuned models have learnt all the generic transformations from multiple speakers' data and also the subject specific transformations in both N2F and N2S cases. In addition to reporting improvements using AstNet over the baseline [1], we predict the duration of the  $rate_{out}$ articulatory trajectory. The relative percentage improvements of AstNet finetune over baseline [1] for all subjects (F1, F2, F3, M1, M2) are 2.09, 1.79, 2.75, 4.32 and 2.84\% respectively and for N2F transformation 5.74, 7.33, 3.17, 6.36 and 10.33\% respectively for N2S transformation.
Despite the fact that the architecture complexity of AstNet is much more than the models used in \cite{astha} and the corresponding relative percentage improvements are all under 11\%, it is still justifiable that AstNet is more suitable  for rate transformation of articulatory movements as it also predicts the duration of the target articulatory movements at $rate_{out}$.
\vspace{-0.3cm}
\subsection{Analysis on the range of articulatory movements:}

Due to the variation in the speaking rate, the range of movement of articulators is affected. 
For example at fast speaking rate, the articulators might undergo lingual undershoot thus impacting there displacement \cite{felge,gay1981} and similarly at slower rate due to hyper-articulation articulatory movements for some phones peak. To verify to what extent AstNet has learnt amplitude scaling, we compute the standard deviation of articulatory trajectories (SDAT) for each sentence \cite{ILLA202075}.
 Figure \ref{amp analysis} report the analysis on characterizing range of movements of articulators which are transformed using AstNet in comparison with input neutral and original trajectories at target speaking rate.
In Figure \ref{amp analysis} we perform amplitude analysis over all 18 articulators for both N2F (top subplot) and N2S (bottom subplot) transformations. For each articulator, we box-plot SDAT over 460 utterances original trajectories at neutral rate as well as original and predicted ones at fast/slow rates. 
For most of the articulators in N2F, we observe there is a reduction in SDAT in fast rate compared to neutral rate, when original articulators are considered. The same trend of reduction in SDAT is also observed for transformed articulatory trajectories. Similar trends are observed in N2S, SDAT increases from neutral to slow rate for both original and transformed trajectories.

\subsection{Phone specific duration analysis:}
  Figure \ref{duration_analysis_n2f} illustrates phoneme duration for N2F and N2S transformations in top and bottom subplots, respectively. We have performed this analysis over a set of 50 phonemes and for each phoneme we box-plot the duration of all utterances for neutral, fast and the predicted fast articulatory movements. We have rank ordered (ascending) them in terms of the absolute median different between the duration at source neutral rate and target fast/slow rates. The durations of the top 15 and bottom 15 from this ranked ordered phonemes are illustrated in Figure \ref{duration_analysis_n2f}.
We observe in Figure \ref{duration_analysis_n2f} (top plot) that, for all phones, neutral duration values are higher than both fast and fast-predicted, as time taken to speak any phoneme at the neutral rate will be longer than that at the fast rate. It is also observed that the durations of the original and transformed articulatory trajectories are similar. A phoneme specific paired t-test between the original durations and the ones predicted by N2F AstNet shows that the durations of the original and transformed articulatory trajectories are significantly ($p<$0.01) different only for seven phonemes, namely 'AX', 'R', 'DD', 'TD', 'T', 'EY', 'AY'.This suggests that AstNet is able to learn phone level duration transformations. This is also true with N2S AstNet, where we observe an increase in the duration from neutral to slow speaking rate and the duration of the original and transformed trajecotories are significantly ($p<$0.01) different only for one phoneme, namely 'AO'.
\vspace{-0.2cm}
\section{Conclusions}
In this work, we have modeled transformation of articulatory movements (N2F and N2S) using encoder-decoder framework with attention network (AstNet), which outperforms the baseline \cite{astha}. Using this model not only do we estimate articulatory movement trajectory from one to another speaking rate, but also predict the duration of the resulting trajectory thus eliminating the need of pre-processing time-alignment of trajectories. We analyze phoneme specific transformations among articulatory movements corresponding to different speaking rates, as it is reported that different sound units are affected differently in different speaking rates \cite{kuwabara1997acoustic}. We also perform an amplitude analysis of predicted and ground-truth trajectories using SDAT and validating that apart from learning time-scaling, model also learns amplitude-scaling.

\vspace{-0.3cm}
\noindent\rule{4cm}{0.4pt}\\
{\footnotesize
Authors thank all the subjects for their participation in the data collection. We also thank the Department of Science and Technology, Govt. of India for their support in this work.}

\bibliographystyle{IEEEtran}

\bibliography{mybib}
\end{document}